\documentstyle[a4]{article}

\newcommand{\beqn}{\begin{equation}}
\newcommand{\eeqn}{\end{equation}}
\newcommand{\benr}
    {\begin{enumerate}
     \def\theenumi{\roman{enumi}}
     \def\labelenumi{\theenumi)}}
\newcommand{\benuma}
    {\begin{enumerate}
     \def\theenumi{\alph{enumi}}
     \def\labelenumi{\theenumi)}}
\newcommand{\benumi}
    {\begin{enumerate}
     \def\theenumi{\roman{enumi}}
     \def\labelenumi{\theenumi)}}
\newcommand{\eenr}{\end{enumerate}}

\begin{document}

\begin{titlepage}

\title{QUANTUM LATTICE SOLITONS\thanks{hep-tt/9406147, submitted to
Physica D}}
\author{ A.C. Scott \thanks{Laboratory of Applied Mathematical Physics,
Technical University of Denmark, DK-2800 Lyngby, Denmark,  and
Department of Mathematics, University of Arizona, Tucson AZ 85721 USA
}   \\ J.C. Eilbeck  \thanks{ Department of Mathematics, Heriot-Watt
University, Riccarton, Edinburgh EH14 4AS, UK} \\ and \\
 H. Gilh\o j \thanks{Department of Physical Chemistry, Technical
University of Denmark, DK-2800 Lyngby, Denmark } }

\maketitle
\end{titlepage}

\begin{abstract}

The number state method is used to study soliton bands for three
anharmonic quantum lattices:
i) The discrete nonlinear Schr\"{o}dinger equation, ii) The
Ablowitz-Ladik system, and iii) A fermionic polaron model.  Each of
these systems is assumed to have $f$-fold translational symmetry in one
spatial dimension, where $f$ is the number of freedoms (lattice
points).  At the second quantum level $(n=2)$ we calculate exact
eigenfunctions and energies of pure quantum states, from which we
determine binding energy $(E_{\rm b})$, effective mass $(m^{*})$ and
maximum group velocity $(V_{\rm m})$ of the soliton bands as functions
of the anharmonicity in the limit $f \rightarrow \infty$.  For
arbitrary values of $n$ we have asymptotic expressions for $E_{\rm b}$,
$m^{*}$, and $V_{\rm m}$ as functions of the anharmonicity in the
limits of large and small anharmonicity.  Using these expressions we
discuss and describe wave packets of pure eigenstates that correspond
to classical solitons.

\end{abstract}

\section{Introduction}

Throughout the development of modern nonlinear dynamics the
investigation of lattices has played a significant role.  Often the
motivation for such studies is that molecular crystals are lattices,
and in these applications quantum effects cannot be ignored.  Typical
experiments---such as infra-red absorption, Raman scattering, and
neutron diffraction---deal with line spectra, where every line
corresponds to a pair of quantum states, each with a particular number
of quanta.  At larger quantum numbers---approaching the correspondence
limit---one is often interested in knowing how quantum corrections
alter the results of classical calculations.  Our aim in this paper is
to present some results of an exact theory of lattice solitons in the
quantum regime.  To this end we consider the following specific models,
which are defined on one-dimensional lattices of $f$ freedoms (or
lattice sites) with periodic boundary conditions and with $\gamma$
being the ratio of anharmonicity to nearest neighbor hopping energy.

\vspace{.2in}
{\bf (i) The quantum discrete nonlinear Schr\"{o}dinger (QDNLS)
equation}
\vspace{.2in}

This system arises in the study of molecular vibrations in
one-dimensional chains such as benzene and certain molecular crystals
\cite{els85,se86a,se86b,sbe89,bes90}.  With site energies scaled out
 through a gauge transformation the
reduced Hamiltonian operator is
\begin{equation}
\hat H_1 = -\sum\sp f_{j=1} \left[b_j\sp {\dagger}b_{j+1} + b_j\sp
{\dagger}b_{j-1}+{\gamma \over 2} b_j\sp {\dagger} b_j\sp {\dagger} b
_j b_j\right],\label{h1}
\end{equation}
where $b_j\sp {\dagger}$ and $b_j$ are standard bosonic raising and
lowering operators satisfying the commutation relations 
$[b_i,b_j] = [b_i\sp {\dagger},b_j\sp
{\dagger}]=0$, $[b_i,b_j\sp
{\dagger}]=\delta_{ij} $ at each freedom.

\vspace{.2in}
{\bf (ii) The quantum Ablowitz-Ladik (QAL) equation }
\vspace{.2in}

The Ablowitz-Ladik equation is of interest because the corresponding
classical system is integrable via the inverse scattering method
\cite{al76b} and the quantum system is a simple example of a $q$-boson
model \cite{ma89}.  The reduced Hamiltonian  operator is
\cite{ku81,gik84}
\begin{equation}
\hat H_2 = - \sum_{j=1}\sp f \left[ b_j\sp {\dagger}( b_{j+1}+ b_{j-1})
\right],\label{h2}
\end{equation}
where $ b_j\sp {\dagger}$ and $ b_j$ are operators satisfying
``$q$-deformed'' commutation relations
\[
[ b_j\sp {\dagger}, b_k\sp {\dagger}]=[ b_j, b_k]=0,\quad
[ b_j, b_k\sp {\dagger}]= \left(1+{\gamma \over 2} b_j\sp {\dagger}
b_k\right) \delta_{jk}\,.
\]

\vspace{.2in}
{\bf (iii) A fermionic polaron (FP) model }
\vspace{.2in}

This model describes the dynamics of electrons in a one-dimensional
crystal \cite{mf84} and is related to the XXZ spin chain model 
\cite{pz86}.  The reduced
Hamiltonian operator is \begin{equation}
\hat H_3 = - \sum\sp f_{j=1}\left[a_j\sp {\dagger} a_{j+1}
+ a_j\sp {\dagger} a_{j-1} + \gamma a_j\sp {\dagger} a_j a_{j+1}\sp
{\dagger} a_{j+1}\right],\label{h3}
\end{equation}
where $a_j\sp {\dagger}$ and $ a_j$ are standard fermion raising and
lowering operators satisfying the anticommutation
relations $\{a_i,a_j\} = \{a_i\sp {\dagger},a_j\sp {\dagger}\}=0$,
$\{a_i,a_j\sp
{\dagger}\}=\delta_{ij} $.

\vspace{.2in}

These systems are chosen to exhibit three different types of
fundamental quanta: i) standard bosons, ii) $q$-deformed bosons, and
iii) standard fermions.  A classical analog (or correspondence limit)
exists only in the first two examples.

Since each of these three models is assumed to satisfy periodic
boundary conditions, the Hamiltonians are invariant under the action of
the translation operator with eigenvalue ${\rm exp}\,({\rm i}k)$, where
$k$ is the crystal momentum.  For a fixed number of quanta $(n)$ the
energy eigenstates for a particular value of $k$ are separated into
bands, and the band of lowest energy is of the form \cite{bes90}
\begin{equation}
E = E_n(k)\;, \label{a5}
\end{equation}
which is of particular interest for two reasons:
\begin{itemize}

\item Within this band there is but a single eigenstate for each value
of $k$.

\item The $n$ quanta of the eigenfunctions for the band are located
more closely together than in the bands of higher energy.

\end{itemize}

We call this lowest band the {\it soliton band} and use Equation
(\ref{a5}) to compute the binding energy, effective mass, and maximum
group velocity of the corresponding quantum soliton.

In the following section we sketch our procedure for computing $E_n(k)$
as a function of the ratio of anharmonicity to nearest neighbor hopping
energy ($\gamma$), and in Section 3 we give complete descriptions of
the bands of our three models for the second quantum level ($n=2$).
Asymptotic expressions for $E_n(k)$ at small and large values of
$\gamma$ are derived in Section 4.  In Section 5 we use these results
to construct wave packets---over $n$ and $k$---of pure eigenstates that
correspond to classical solitons, and some conclusions are presented in
Section 6.  Throughout the discussion we assume units of energy and
time for which $\hbar =1$.

We have chosen to call our method of analysis the number state method,
to distinguish it from the Quantum Inverse Scattering Method
\cite{ko93,ki82}.  The advantages and disadvantages of each of these
two methods is discussed at some length in \cite{esse92}.  In all cases
where both techniques have been used they give (as expected) the same
results.  However it is important to note that the first model we
consider (QDNLS) is nonintegrable and cannot be analysed by the QISM.

\section{The number state method of analysis}
\setcounter{equation}{0}

In our study of these three quantum lattices we take advantage of the
fact that each Hamiltonian operator commutes with a number operator
$\hat N$, which counts the number of fundamental quanta (bosons,
$q$-deformed bosons, or fermions) associated with each degree of
freedom \cite{se86a,esse92}.  Thus a general eigenfunction of $\hat N$
can be written in the form  \begin{equation} |\psi_n\rangle =
\sum_{l=1}^p c_l |\phi_l\rangle\;, \label{b1} \end{equation} where $p =
p(n,f)$ is the number of different ways that $n$ quanta can be arranged
on $f$ freedoms, $|\phi_l\rangle$ is the number state corresponding to
a particular arrangement, and the $\{c_l\}$ are a set of $p$ arbitrary
complex constants.  For example if we are placing two bosons $(n=2)$ on
two freedoms $(f=2)$, there are three possible number states $(p=3)$:

$|\phi_1\rangle = |2\rangle|0\rangle$,
$|\phi_2\rangle = |1\rangle |1\rangle$, 
and 
$|\phi_3\rangle = |0\rangle |2\rangle$,
which for typographical convenience we write as [20], [11], and [02].
Requiring $|\psi_n\rangle$ to satisfy the time independent
Schr\"{o}dinger equation
\begin{equation}
\hat H |\psi_n\rangle = E |\psi_n\rangle \label{b2}
\end{equation}
generates a $p \times p$ matrix equation for the $\{c_l\}$'s, which
determines the energy eigenvalues, and the eigenvectors---upon
substitution into Equation (\ref{b1})---yield the corresponding
stationary state wave functions.  We refer to this technique as the
``number state method'' (NSM) to differentiate it from the quantum
inverse scattering method (QISM) over which it has some theoretical and
computational advantages \cite{esse92}.

In effect the NSM diagonalizes the infinite Hamiltonian matrix into
finite blocks of size $p$.  For the bosons of $\hat H_1$ and the
q-deformed bosons of $\hat H_2$
\begin{equation}
p_1 = p_2 = {(n+f-1) ! \over n!(f-1)!} \;. \label{b3}
\end{equation}
Only one of the fermions of $\hat H_3$ can be placed on a single
freedom so in this case
\begin{equation}
p_3 = { f! \over n!(f-n)!} \;. \label{b4}
\end{equation}
A moment's reflection will convince the reader that $p$ can easily
become inconveniently large, but this is not a  problem particular to
the number state method.  Exact quantum wave functions on lattices are
complicated objects, and the same difficulty appears with the QISM
\cite{esse92}.  Each of our three systems has translational symmetry so
these $p\times p$ blocks can be further diagonalized into smaller
blocks with fixed values of crystal momentum $k$, where $\tau = {\rm
exp}\,({\rm i}k)$ is an eigenvalue of the translation operator $\hat
T$.  $\hat{T}$ is defined by the property $\hat T b\sp \dagger_j =b\sp
\dagger_{j+1} \hat T$ so that $\hat T[n_1 n_2\ldots n_f] = [n_f
n_1\ldots n_{f-1}]$.
  
In this manner we find $E = E(k)  $ for every allowed momentum state
with a minimum of computational effort.

\vspace{.2in}

As a simple example of this method, consider the first quantum level
$(n=1)$.  The energy bands for all three models are identical because
there are just $f$ ways that a single quantum can be placed on $f$
freedoms.  Thus we find
\begin{equation}
|\psi_1(k)\rangle = {1 \over \sqrt{f}}\sum_{j=1}^f \left(
{\rm e}^{{\rm i}k}
\hat{T}\right)^{j-1}[100\cdots0]\,, \label{ps1}
\end{equation}
and
\begin{equation} 
E_1(k) = -2{\rm cos}\,k \;, \label{b7}
\end{equation}
where $k=2\pi \nu / f$ and $\nu = 0, \pm1, \pm2, \ldots, 
\pm(f/2 - 1), f/2$ for $f$ even and
$0, \pm1, \pm2, \ldots, \pm(f-1)/2$ for $f$ odd.

Since the effective mass $(m^{*})$ is defined as
\begin{equation}
E_1(k) =  E_1(0) + {k\sp 2 \over 2m\sp * } + {\rm
O}(k\sp 4)\,,  \label{b8}
\end{equation}
we have 
\begin{equation}
m^* = 1/2 \,. \label{b10}
\end{equation}

\section{The second quantum level}
\setcounter{equation}{0} 

Here we show how to compute the exact eigenstates and energy
eigenvalues of the soliton bands at the second quantum level $(n=2)$.
To this end we display number operators, $\hat N_i\,,\,\,i=1,2,3$ that
commute with each other and the Hamiltonian, and we construct the most
general eigenfunctions of $\hat N_i$ and the translation operator $\hat
T$ as a sum of products of elementary number states
\cite{egs93,ges93}.

\vspace{.2in}
{\bf (i) QDNLS }
\vspace{.2in}

$\hat H_1$, the Hamiltonian of Equation (\ref{h1}), commutes with the
number operator
\begin{equation}
\hat N_1 = \sum\sp f_{j=1}b_j\sp {\dagger}b_j\;, \label{c1}
\end{equation}
which has eigenvalue $n$.  As discussed in the previous section, the
most general eigenfunction of $\hat N_1$ is a sum of products of
elementary number states of the form $[n_1 n_2\ldots n_f]$, where $n =
n_1 + n_2 + \cdots +n_f.$
  
For $n=2$ and $f$ odd a general eigenfunction of both $\hat N_1$ and
$\hat T$ is
\begin{equation}
\begin{array}{lcl}
|\psi_2\rangle & = &
 {1 \over \sqrt{f}}\{c_1\sum_{j=1}\sp {f} (\tau\hat T)\sp 
{j-1}[20\cdots0]
 + c_2\sum_{j=1}\sp {f}(\tau\hat T)\sp {j-1}[110\cdots0]+\\
&& + c_3\sum_{j=1}\sp {f}(\tau\hat T)\sp {j-1}[1010\cdots0]+\cdots +\\
&& + c_{(f+1)/2}\sum_{j=1}\sp {f}(\tau\hat T)\sp
{j-1}[10\cdots010\cdots00]\}\,,\label{psi1} \end{array} \label{c3}
\end{equation}
where $\tau = {\rm exp}({\rm i}k)$
is the eigenvalue of $\hat T$ that corresponds to the wavenumber
$k = 2\pi{\rm i}\nu/f$.  To ensure that
$ \langle\psi_2|\psi_2\rangle = 1$,
it is necessary that the $c_l$'s be normalized as
\begin{equation}
\sum_{l=1}\sp {(f+1)/2}|c_l|\sp 2 = 1\,.\label{norm1}
\end{equation}
Requiring that $\hat H_1 |\psi_2\rangle = E|\psi_2\rangle$ leads to the
matrix equation
$Q_1(\tau) {\bf c} = E {\bf c}$,
where ${\bf c} ={\rm col}(c_1,c_2,\ldots,c_{(f+1)/2})$ and
 $Q(\tau)$ is the $[(f+1)/2]\times [(f+1)/2]$
matrix
\begin{equation}
 Q_1(\tau) = -  \left(\matrix{\gamma&q\sp *\sqrt2&{}&{}&{}&{}\cr
q\sqrt2&0&q\sp *{}&{}&{}&{}\cr
{}&q&0&q\sp *{}&{}&{}\cr
{}&{}&\ddots &\ddots &\ddots &{}\cr
{}&{}&{}&q&0&q\sp *\cr
{}&{}&{}&{}&q&p\cr}\right)\,, \label{q1}
\end{equation}
and
\begin{equation}
 q \equiv 1+\tau ,\quad p \equiv (\tau\sp {(f+1)/2} + \tau\sp {(f-1)/2})\,.\label{qp}
\end{equation}

For $f$ even (which is a special case of the condition $f \bmod n =
0$), the wave function depends on whether the integer $\nu$ in the
translational eigenvalue $\tau = {\rm exp}(2\pi{\rm i}\nu/f)$ is even
or odd.  For $\nu$ even
\begin{equation}
\begin{array}{lcl}
|\psi_2\rangle & = &
 { 1 \over \sqrt{f}}\{c_1\sum_{j=1}\sp {f} (\tau\hat T)\sp 
{j-1}[20\cdots0]
 + c_2\sum_{j=1}\sp {f}(\tau\hat T)\sp {j-1}[110\cdots0]+\\
&& + c_3\sum_{j=1}\sp {f}(\tau\hat T)\sp {j-1}[1010\cdots0]+\cdots +\\
&& + (c_{(f+1)/2}/\sqrt{2})\sum_{j=1}\sp {f}(\tau\hat T)\sp
{j-1}[10\cdots010\cdots00]\}\,, \end{array} \label{cc3}
\end{equation}
whereas for $\nu$ odd the last term in the sum is omitted.

For $f$ and $\nu$ both even, $Q_1(\tau)$ is the $(f/2 + 1)\times(f/2+1)$ 
matrix
\begin{equation}
 Q_1(\tau) = -  \left(\matrix{\gamma&q\sp *\sqrt2&{}&{}&{}&{}\cr
q\sqrt2&0&q\sp *{}&{}&{}&{}\cr
{}&q&0&q\sp *{}&{}&{}\cr
{}&{}&\ddots &\ddots &\ddots &{}\cr
{}&{}&{}&q&0&q\sp *\sqrt{2}\cr
{}&{}&{}&{}&q\sqrt{2}&0\cr}\right)\,, \label{qq1}
\end{equation}
where $q$ is defined in Equation (\ref{qp}).  For odd $\nu$,
$Q_1(\tau)$  is the $(f/2)\times(f/2)$ matrix obtained from Equation
(\ref{qq1}) by omitting the last row and column.

Finding the eigenvalues of tridiagonal matrices of this sort is an
interesting exercise in analysis: full details are given \cite{ep93}.
Exact results can be obtained in the limit $f \rightarrow \infty$; for
finite but large $f$ analtic correction factors can be found.

In the limit $f \rightarrow \infty$ the soliton band
has the energy 
\begin{equation} E_2(k) = -\sqrt{\gamma\sp 2 + 16{\rm cos}\sp 2\,
(k/2)} \;.
\label{cc1} 
\end{equation}
Defining the binding energy, $E_{\rm b}$, as the difference between the
energy of the soliton band at $k=0$ and the bottom of the continuum
band, one finds that
\begin{equation}
E_{\rm b} = \sqrt{\gamma\sp 2 + 16} - 4 \;, \label{ccc2}
\end{equation}
and with effective mass defined as in Equation (\ref{b8}) 
\begin{equation}
m\sp * =  { \sqrt{\gamma\sp 2 + 16} \over 4}     \;, \label{ccc3}
\end{equation}
and maximum group velocity
\begin{equation}
V_{\rm m} \equiv \left[{dE \over dk}\right]_{k=\pi/2} =  { 4 \over \sqrt{\gamma\sp 2 + 8}}     \;.
\label{cc4} \end{equation}

In Figure 1(i) we display the energy eigenvalues calculated from
Equation (\ref{q1}) with $\gamma = 3$ in the limit $f \rightarrow
\infty$.  The existence of a quasi-continuum and  a lower discrete band
is clear.

\begin{figure}
\vskip 17 cm
\includegraphics{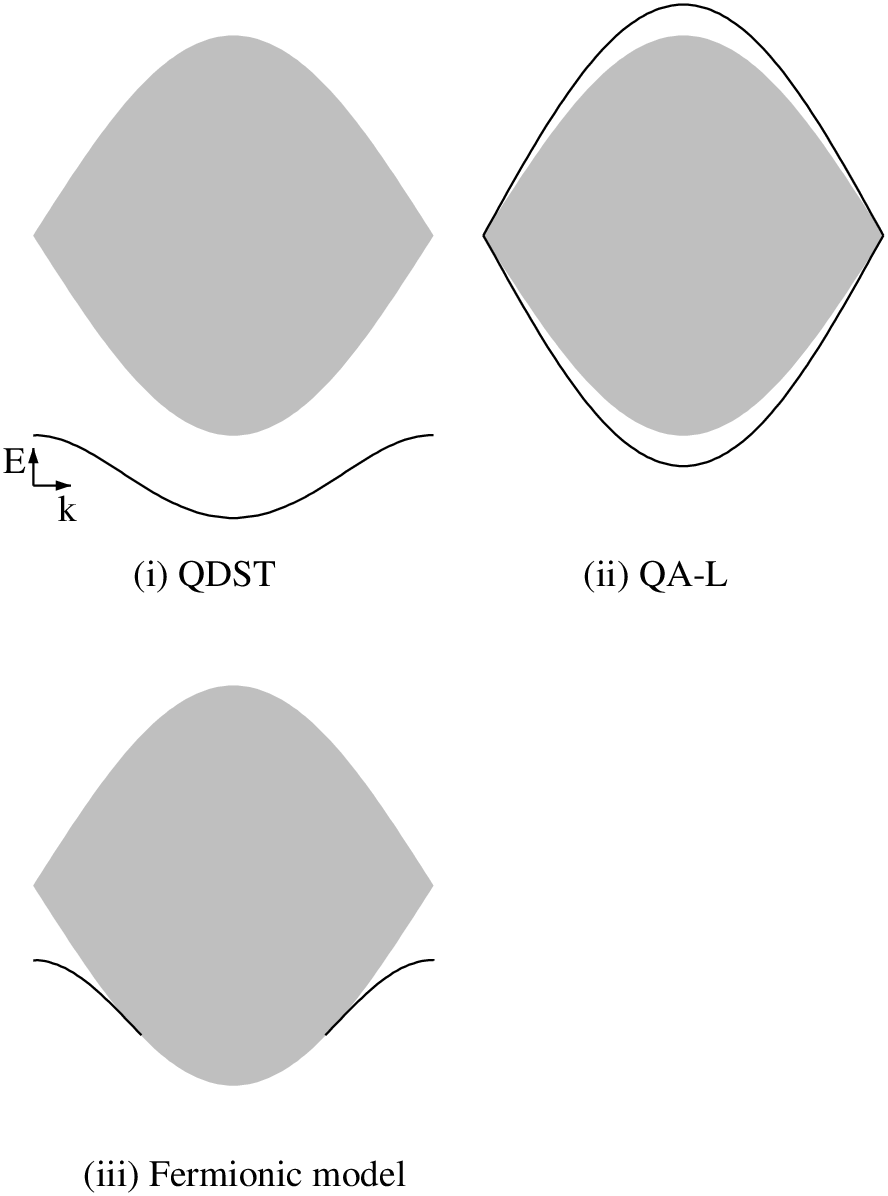}
\caption{Soliton bands in the three models considered in the text}
\end{figure}

\vspace{.2in}
{\bf (ii) QAL }
\vspace{.2in}

The Hamiltonian $\hat H_2$ commutes with the translation operator
 and with the number operator \cite{sa92a,sa92b}
\begin{equation}
\hat N_2=\sum_{j=1}\sp f {\ln \left(1+{\gamma \over 2}  b_j\sp 
{\dagger} b_j\right)\over
{\ln \left(1+{\gamma \over 2}\right) }}. \label{c4}
\end{equation}

Taking the eigenvalue $n$ of $\hat N_2$ to be 2 the wave functions are
as in Equations (\ref{c3}) and (\ref{cc3}).  For $f$ odd the energy
eigenvalues are found from the $[(f+1)/2] \times [(f+1)/2]$ matrix
\begin{equation}
 Q(\tau) = - \left(\matrix{0&q\sp *\sqrt{2+\gamma/2}&{}&{}&{}&{}\cr
q\sqrt{2+\gamma/2}&0&q\sp *{}&{}&{}&{}\cr
{}&q&0&q\sp *{}&{}&{}\cr
{}&{}&\ddots &\ddots &\ddots &{}\cr
{}&{}&{}&q&0&q\sp *\cr
{}&{}&{}&{}&q&p\cr}\right)\,,\label{q2}
\end{equation}
where $q, p$ are as defined in (\ref{qp}).    

For $f$ and $\nu$ both even, $Q_2(\tau)$ is the $(f/2 + 1)\times
(f/2 + 1)$ matrix
\begin{equation}
 Q(\tau) = - \left(\matrix{0&q\sp *\sqrt{2+\gamma/2}&{}&{}&{}&{}\cr
q\sqrt{2+\gamma/2}&0&q\sp *{}&{}&{}&{}\cr
{}&q&0&q\sp *{}&{}&{}\cr
{}&{}&\ddots &\ddots &\ddots &{}\cr
{}&{}&{}&q&0&q\sp*\cr
{}&{}&{}&{}& q &0\cr}\right)\,,\label{qq2}
\end{equation}
whereas for $\nu$ odd it is the $(f/2)\times(f/2)$ matrix obtained from
Equation (\ref{qq2}) by omitting the final row and column.

In the QAL case we find two soliton bands, which are shown in Figure
1(ii).  The top band corresponds to the classical AL soliton that
alternates in sign between each lattice point.

In the limit $f \rightarrow \infty$ the soliton band has energy
\cite{ep93}
\begin{equation}
E_2(k) = \pm { 2{\rm cos}\,(k/2)(\gamma+4) \over \sqrt{2\gamma + 4}} \;. \label{cc5}
\end{equation}
Defining the binding energy, $E_{\rm b}$, as above, one finds that
\begin{equation}
E_{\rm b} = { 2(\gamma+4) \over \sqrt{2\gamma + 4}} - 4 \;, \label{cc6}
\end{equation}
and the effective masses are 
\begin{equation}
m\sp * =  \pm { 2\sqrt{2\gamma + 4} \over \gamma + 4}     \;, \label{cc7}
\end{equation}
where the ``$+$'' corresponds to the lower band and the ``$-$'' sign to
the upper band.  In this case the maximum group velocity is
\begin{equation}
V_{\rm m} \equiv \left[{dE \over dk}\right]_{k=\pi} =  {\gamma + 4
\over \sqrt{2\gamma + 4}}     \;.
\label{cc8} \end{equation}

\vspace{.2in}
{\bf (iii) FP }
\vspace{.2in}

The Hamiltonian $\hat H_3$ commutes with the number operator

\begin{equation}
\hat N_3 = \sum\sp f_{j=1}a_j\sp {\dagger} a_j \,. \label{c5}
\end{equation} 

Because of the anti-commutation relations, the order in which one
inserts the fermions into the chain is important \cite{di58}.  We
define the normal ordering to be that in which the fermions are
inserted from left to right.  Hence $a_2\sp {\dagger}a_1\sp
{\dagger}[00] = [11]$,  but $a_1\sp {\dagger}a_2\sp {\dagger}[00] =
-[11]$.  The translation operator is defined by $\hat T a_j\sp
{\dagger} = a_{j+1}\sp {\dagger} \hat T$, so, for example, $\hat T
[1001] = -[1100].$  As was pointed out by Dirac, this makes hand
calculations tedious, but fortunately an algebraic manipulation system
like Mathematica \cite{wo91} can be programmed to take such sign
changes into account.

Again assuming the eigenvalue of $\hat N_3$ to be 2 and $f$ odd, a
general eigenfunction of $\hat N_3$ and $\hat T$ is
\begin{equation}
\begin{array}{r}
|\psi_2\rangle = {1 \over \sqrt{f}}\{c_1 \sum_{j=1}\sp 
f(\tau\hat T)\sp {j-1}[110 \cdots 0] +
c_2 \sum_{j=1}\sp f(\tau\hat
T)\sp {j-1}[1010 \cdots 0] + \\
+ \cdots + c_{(f-1)/2} \sum_{j=1}\sp f(\tau\hat T)\sp {j-1}
[10 \cdots 010
\cdots 00]\}\;.\label{psi3}
\end{array}
\end{equation}

Note that this wavefunction has one less term than that given in
Equation (\ref{psi1}) because no more than one fermion can be assigned
to any one freedom.  Requiring $\hat H_3 |\psi_3\rangle =
E|\psi_3\rangle$  leads to the matrix equation
$ Q_3(\tau) {\bf c} = E {\bf c},
$ where $Q_3(\tau)$ is the $[(f-1)/2]\times [(f-1)/2]$
matrix
\begin{equation} Q_3(\tau) =  -  \left(
\matrix{\gamma&q\sp *&{}&{}&{}&{}\cr
q&0&q\sp *{}&{}&{}&{}\cr
{}&q&0&q\sp *{}&{}&{}\cr
{}&{}&\ddots &\ddots &\ddots &{}\cr
{}&{}&{}&q&0&q\sp *\cr
{}&{}&{}&{}&q&-p\cr}\right)\,,\label{q3}
\end{equation}
and $q,p$ are as defined in (\ref{qp}).

If $\gamma>2$, the resulting eigenvalue plot appears as in Figure 1(i),
but for $0<\gamma<2$, the central part of the soliton band merges
with the quasi-continuum as is shown in Figure (iii).

For $f$ and $\nu$ both even,
\begin{equation}
\begin{array}{r}
|\psi_2\rangle = { 1 \over \sqrt{f}}\{c_1\sum_{j=1}\sp f(\tau\hat T)\sp {j-1}[110 \cdots
0] + c_2 \sum_{j=1}\sp f(\tau\hat
T)\sp {j-1}[1010 \cdots 0] + \\
+ \cdots + (c_{(f-1)/2}/\sqrt{2}) \sum_{j=1}\sp f(\tau\hat T)\sp {j-1}
[10 \cdots 010
\cdots 00]\}\;.\label{pp3}
\end{array}
\end{equation}
and $Q_3(\tau)$ is the $(f/2 -1)\times(f/2 - 1)$ matrix
\begin{equation} Q_3(\tau) =  -  \left(
\matrix{\gamma&q\sp *&{}&{}&{}&{}\cr
q&0&q\sp *{}&{}&{}&{}\cr
{}&q&0&q\sp *{}&{}&{}\cr
{}&{}&\ddots &\ddots &\ddots &{}\cr
{}&{}&{}&q&0&q\sp*\cr
{}&{}&{}&{}&q&0\cr}\right)\,.\label{qq3}
\end{equation}
For $\nu$ odd $|\psi_2\rangle$ is as in Equation (\ref{pp3}) with the
last term omitted, and $Q_3(\tau)$ is as in Equation (\ref{qq3}) but
with the last row and column omitted.

In the limit $f \rightarrow \infty$ the soliton band has energy
\cite{ep93}
\begin{equation}
E(k) = - \left[\gamma + {4 \over \gamma}{\rm cos}\sp 2\left({k \over 2}\right)\right] \;\;\; {\rm
for} \;\;\;\gamma>2{\rm cos}\,(k/2). \label{cc9} 
\end{equation}
For $\gamma > 2$ the binding energy and effective mass are
\begin{equation}
E_{\rm b} = \gamma + { 4 \over \gamma}\;, \label{cc10}
\end{equation}
\begin{equation}
m\sp * =  {\gamma \over 2}     \;, \label{cc11}
\end{equation}
and the maximum group velocity is
\begin{equation}
V_{\rm m} \equiv \left[{dE \over dk}\right]_{k=\pi/2} =  
{2 \over \gamma}\,.   \label{cc12}
\end{equation}

\section{Asymptotic expressions for arbitrary quantum levels}
\setcounter{equation}{0}

Proceeding as in the previous section it is possible---in
principle---to construct block diagonalized Hamiltonian matrices for
any value of the quantum number $n$.  As was noted in Section 2,
however, the sizes of these blocks may grow beyond the limits of
computational convenience (or even possibility) so it is of interest to
consider approximate calculations that are useful in asymptotic
limits.  In this section we present results for arbitrary values of the
quantum level that are asymptotically correct for small or large values
of the anharmonicity parameter.

\subsection{Small $\gamma$ }

Binding energy has been studied in detail for the QDNLS and QAL systems
in reference \cite{msce91}, and with $\gamma \ll 1$ these results are
identical for QDNLS and QAL.  To calculate the binding energy in this
limit it is necessary to know whether the size of the classical soliton
is large or small compared with the number of freedoms.  If $1\gg
\gamma >24/((n+1)f)$, the size of the classical soliton is smaller than
$f$ \cite{msce91} and both the QDNLS and the QAL equations are well
approximated by the continuum nonlinear Schr\"{o}dinger equation for
which the binding energy of a quantum soliton is \cite{kk76}
\begin{equation}
E_{\rm b} = {\gamma\sp 2 \over 48}n(n\sp 2-1)\,.
\end{equation}

To calculate the effective masses for QDNLS and QAL in the limit of
small $\gamma$ we refer to Equation (\ref {b7}), which gives the energy
as a function of the wave number for a single quantum.  Since the
classical problem is linear in this limit, the energy for $n$ quanta
with wave numbers: $k_1, k_2, \ldots, k_n$ is just the sum:
$-2\sum_j{\cos}\,k_j$.  For the $n$-quantum wave function $k =
\sum_jk_j$, and the lowest value of energy is found for $k_1 = k_2 =
\cdots = k_n$; thus in this limit
\begin{equation}
E_n(k) = -2n{\cos}\,(k/n)
\end{equation}
so
\begin{equation}
m\sp {*} = {n \over 2}\,,
\end{equation}
and the maximum group velocity of a wave packet is
\begin{equation}
V_{\rm m} \equiv \left[{dE \over dk}\right]_{k=\pi/2} = 2{\rm sin}\,
\left({\pi \over 2n}\right)\,. 
\end{equation}

For the FP system the soliton band merges into the continuum at small
$\gamma$ so $E_{\rm b}$ and $m\sp *$ are not defined in this limit.

\subsection{Large $\gamma$}

In the limit of large $\gamma$ it is evident that for $n=2$ the
dominant elements of the matrices $Q_1$, $Q_2$, and $Q_3$, which are
displayed in Equations (\ref{q1}), (\ref{q2}), and (\ref{q3}), are the
$2\times 2$ submatrices in the upper left hand corners.  For arbitrary
values of $n$, perturbation theory in $\gamma\sp {-1}$ shows that to
calculate the leading $k$-dependent terms it is only necessary to
consider the sequence of interactions
\begin{equation}
[n] \leftrightarrow [n-1,1] \leftrightarrow [n-2,2] \leftrightarrow 
\cdots 
\leftrightarrow [2,n-2] \leftrightarrow [1,n-1] \leftrightarrow [n]\,. \label{kcycle}
\end{equation}
For the QDNLS and QAL systems this sequence requires an approximate wave function of the form
\begin{equation}
\begin{array}{lcl}
|\psi_n\rangle & \doteq &
 { 1 \over \sqrt{f}}\{c_1\sum_{j=1}\sp {f} (\tau\hat T)\sp {j-1}
[n0\cdots0]
 + c_2\sum_{j=1}\sp {f}(\tau\hat T)\sp {j-1}[(n-1)10\cdots0]+\\
&& + c_3\sum_{j=1}\sp {f}(\tau\hat T)\sp {j-1}[(n-1)00\cdots1]+
\cdots +\\
&& + c_n\sum_{j=1}\sp {f}(\tau\hat T)\sp
{j-1}[(n/2)(n/2)0\cdots0]\}\,, \end{array} \label{psie}
\end{equation}
with $n$ even, and 
\begin{equation}
\begin{array}{lcl}
|\psi_n\rangle & \doteq &
 { 1 \over \sqrt{f}}\{c_1\sum_{j=1}\sp {f} (\tau\hat T)\sp {j-1}
[n0\cdots0]
 + c_2\sum_{j=1}\sp {f}(\tau\hat T)\sp {j-1}[(n-1)10\cdots0]+\\
&& + c_3\sum_{j=1}\sp {f}(\tau\hat T)\sp {j-1}[(n-1)00\cdots1]+
\cdots +\\
&& + c_{n-1}\sum_{j=1}\sp {f}(\tau\hat T)\sp
{j-1}[((n+1)/2)((n-1)/2)0\cdots0]\,,+\\
&& + c_n\sum_{j=1}\sp {f}(\tau\hat T)\sp
{j-1}[((n+1)/2)0\cdots0((n-1)/2)]\}\, \end{array} \label{psio}
\end{equation}
for n odd.

\vspace{.2in}
{\bf (i) QDNLS }
\vspace{.2in}

As an illustrative example, consider the QDNLS equation with $n$ even
for which the $n\times n$ matrix $-\tilde{Q}_1(\tau)$ formed from the
approximate wave function in Equation (\ref{psie}) is
\begin{equation}
\left(\matrix{\gamma {n(n-1)\over 2}&\sqrt{n}&\sqrt{n}&{}&{}\cr
\sqrt{n}&\gamma {(n-1)(n-2) \over 2}&0&{}&{}\cr
\sqrt{n}&0&\gamma{(n-1)(n-2) \over 2}&\sqrt{3(n-2)}&{}\cr
{}&\ddots&\ddots&\ddots &\ddots \cr
{}&{}&0&\gamma({n\sp 2 \over 4} -{n \over 2}+1)&
\tau\sp {-1}\sqrt{{n \over 2}({n \over 2}+1)}\cr 
{}&{}&\sqrt{{n \over 2}({n \over 2}+1)}&
\tau\sqrt{{n \over 2}({n \over 2}+1)}&\gamma({n\sp 2 \over 4}-{n\over 2})\cr}\right)\,, 
\end{equation}

We have no exact results for the eigenvalues of such approximate
matrices for arbitrary values of $n$.  However we have done extensive
investigations of the asymptotic expansions of the eigenvalues in
powers of $1/\gamma$ for finite values of $n$ using Mathematica
\cite{wo91} and perturbation theory.  Clearly the most negative
eigenvalue has a leading term $-n(n-1)/2\gamma$.  Higher order terms
can be calculated but they are of secondary interest except for the
lowest order term that depends on $k$.  Perturbation theory shows that
this term is of order $1/\gamma\sp {n-1}$.  In summary, the most
negative root of the determinental equation, ${\rm det}\,[\tilde{Q}_1 -
IE]=0$, defines a soliton band of the form
\begin{equation}
E_n(k) \doteq -{1 \over 2}n(n-1)\gamma -\left({2n \over (n-1)!\gamma\sp {n-1}}\right){\rm
cos}\,k\,, \label{ek}
\end{equation}
where the symbol ``$\doteq$'' indicates that the first term on the
right hand side is correct to O($\gamma\sp {-1}$) while the second
($k$-dependent) term is correct to O($\gamma\sp {1-n}$).  In other
words, terms of order O($1/\gamma$) that do not depend on $k$ have been
dropped. The second term on the right comes from the perturbation
correction.

Equation (\ref{ek})
implies that for the QDNLS equation 
\begin{equation}
E_{\rm b} \doteq {1 \over 2} n(n-1)\gamma\;,
\end{equation}
\begin{equation}
m\sp * \doteq {(n-1)! \over 2n} \gamma\sp {n-1}\;,
\end{equation}
and
\begin{equation}
V_{\rm m} \equiv \left[{dE \over dk}\right]_{k=\pi/2} \doteq {2n 
\over (n-1)!
\gamma\sp {n-1}}.\label{v1} \end{equation}

\vspace{.2in}
{\bf (ii) QAL }
\vspace{.2in}

For the QAL equation the wave functions are of the forms given in
Equations (\ref{psie}) and (\ref{psio}) except that it is convenient to
permute the order of terms such that
$\{c_{1},c_{2},\dots,c_{n}\}\rightarrow
\{c_{n-3},c_{n-5},\dots,c_3,c_1,c_2,c_4,\dots,c_{n-2},c_n,c_{n-1}\}$.
With this rearrangement the corresponding $n\times n$ matrix is to
leading order in $\gamma\sp {-1}$
\begin{equation}
 \tilde{Q}_2(\tau) = -  \left(\matrix{0&\beta&0&{}&{}&0&\beta\cr
\beta&0&\beta&0&{}&{}&0\cr
0&\beta&0&\beta&0&{}&{}\cr
{}&\ddots&\ddots&\ddots &\ddots &\ddots &{}\cr
{}&{}&\ddots&\ddots&\ddots&\beta&0\cr
0&{}&{}&0&\beta&0&\beta \tau\sp {-1}\cr
\beta&0&{}&{}&0&\beta \tau&0\cr}\right)\,,\label{qqq2}
\end{equation}
where
\begin{equation}
\beta(n) \equiv \left({\gamma \over 2}\right)\sp {(n-1)/2}\,.
\label{beta}
\end{equation}

Now define $Q_n= {\rm det}[\beta\sp {-1}\tilde{Q}_2(\tau)-eI]$. 
Expanding
$Q_n$ by final rows and columns gives eventually that
$$
Q_n=(-1)\sp {n+1}2\cos k-2P_{n-2}+eP_{n-1}
$$
where $P_{n-1}$ is the tridiagonal determinant formed by removing the
final row and column of $Q_n$.

A standard calculation shows that $P_n(e)$ is a polynomial satisfying
the recursion relation
\begin{equation}
P_n = eP_{n-1} - P_{n-2}\,,
\end{equation}
where $P_1 = e$ and $P_2 = e\sp 2 - 2$.  Thus $P_n(e) = 2U_n(e/2)$,
where $U_n(x)$ is a Chebyshev polynomial of the second kind.  Hence
\begin{eqnarray}
Q_n & = &(-1)\sp {n+1} 2 \cos k + U_n - U_{n-2} \cr
  & = &(-1)\sp {n+1} 2 \cos k + 2 T_n({e\over 2}) 
\end{eqnarray}
Where $T_n(x)$ is a Chebyshev polynomial of the first kind.

A further short
calculation shows that the eigenvalues of (\ref{qqq2}) are
\begin{equation}
E_n(k) = -2{\rm cos}\left({k \over n}\right)\left(
{\gamma \over 2}\right)\sp {(n-1)/2}\,. \label{ek2}
\end{equation}
so for $n>1$
\begin{equation}
E_{\rm b} = 2\left({ \gamma \over 2}\right)\sp {(n-1)/2}\;,\label{eb}
\end{equation}
and
\begin{equation}
m\sp * = {n\sp 2 \over 2}\left({2 \over \gamma}
\right)\sp {(n-1)/2}\;.\label{m}
\end{equation}

The maximum group velocity on the QAL soliton band occurs at the band
edge $(k=\pi)$, therefore
\begin{equation}
V_{\rm m} \equiv \left[{dE \over dk}\right]_{k=\pi} = {2 \over n}{\rm sin}\left({\pi \over
n}\right) \left({\gamma \over 2}\right)\sp {(n-1)/2}\,.\label{ve}
\end{equation}

\vspace{.2in}
{\bf (iii) FP }
\vspace{.2in}

For the FP model a translationally invariant wave function in the large
$\gamma$ limit is
\begin{equation}
\begin{array}{lcl}
|\psi_n\rangle & \doteq &
 { 1 \over \sqrt{f}}\{c_1\sum_{j=1}\sp {f} (\tau\hat T)\sp {j-1}[11\cdots10\cdots0]+\\
 &&+ c_2\sum_{j=1}\sp {f}(\tau\hat T)\sp {j-1}[11\cdots1010\cdots0]+\\
 &&+ c_3\sum_{j=1}\sp {f}(\tau\hat T)\sp {j-1}[011\cdots10\cdots01]+\\
&&+\cdots + c_n\sum_{j=1}\sp {f}(\tau\hat T)\sp
{j-1}[\underbrace{11\cdots1}_{n/2}0\underbrace{11\cdots1}_{n/2}0
\cdots0]\}\,, \end{array}
\label{fpps}
 \end{equation}
for $n$ even.  For $n$ odd there will be a corresponding expression
with the last two terms given by
\[
\begin{array}{c}
 c_{n-1}\sum_{j=1}\sp {f} (\tau\hat T)\sp {j-1}[\underbrace{11\cdots1}_{(n-1)/2}0
\underbrace{11\cdots1}_{(n+1)/2}0\cdots0] +
\\
 c_n\sum_{j=1}\sp {f}(\tau\hat T)\sp
{j-1}[\underbrace{11\cdots1}_{(n+1)/2}0
\underbrace{11\cdots1}_{(n-1)/2}0\cdots0]\}\,.
 \end{array} 
\]
Again it is convenient to reorder to get a periodic tridiagonal
matrix.  For $n$ even, take
$\{c_{1},c_{2},\dots,c_{n}\}\rightarrow
\{c_1,c_3,c_5\dots,c_{n/2},c_{n/2-1},c_{n/2-3},\dots,c_2\}$, with 
a similar permutation in the $n$ odd case.  This leads to a
 $n\times n$ matrix $-\tilde{Q}_3(\tau)$ 
\begin{equation}
  \left(\matrix{(n-1)\gamma&\tau\sp {-1}&0&{}&{}&0&1\cr
\tau&(n-2)\gamma&1&0&{}&{}&0\cr
0&1&(n-2)\gamma&1&0&{}&{}\cr
{}&\ddots&\ddots&\ddots &\ddots &\ddots &{}\cr
{}&{}&\ddots&\ddots&\ddots&\ddots&\ddots\cr
0&{}&{}&0&1&(n-2)\gamma&1\cr
1&0&{}&{}&0& 1&(n-2)\gamma\cr}\right)\,.
\end{equation}

The characteristic equation is calculated in a similar way to the QAL
case.  If we define $Q_n={\rm det}[-\tilde{Q}_3(\tau)+(n-2)\gamma-eI]$,
then
$$
Q_n=0 \Rightarrow (\gamma+e)U_{n-1}(e/2)-2U_{n-2}(e/2)-(-1)\sp 
n 2\cos k=0
$$
To solve this polynomial equation for $e$  by a perturbation series 
around $\gamma=\infty$, put $\gamma=\epsilon\sp {-1}$, $z=-e\epsilon$, to
get
$$
(-1)\sp n z\sp {n-1}(z-1)+(-1)\sp {n+1} \epsilon\sp n 2\cos k
+\sum_{m=1}\sp {M} \epsilon\sp {2m}z\sp {2(M-m)}p_m(z) =0
$$
where $M=\lfloor n/2 \rfloor$ and the $p_m$ are linear functions of $z$
with constant coefficients.  A series expansion shows that one root is
$z=1+O(\epsilon\sp 2)$ and the others are $O(\epsilon)$.  If we seek
the lowest order $k-$dependent correction to the $z=1$ root, along the
lines of eqn (\ref{ek}), it is not difficult to show that this occurs
at $O(\epsilon\sp n)$ and is equal to $\epsilon\sp n 2 \cos k$.
Resubstituting this back into the original problem, we have that the
soliton band is given by \begin{equation} E_n(k) \doteq -(n-1)\gamma -
{2 \over \gamma\sp {n-1}}{\rm cos}\,k\,, \label{efp} \end{equation}
where the symbol ``$\doteq$'' has the same meaning as that discussed
following (\ref{ek}).  This implies that
\begin{equation}
E_{\rm b} = (n-1)\gamma\,,
\end{equation}
\begin{equation}
m\sp {*} = {\gamma\sp {n-1} \over 2}\,,\;\;\;{\rm and}\label {mfp}
\end{equation}
\begin{equation}
V_{\rm m} \equiv \left[{dE \over dk}\right]_{k=\pi/2} = 
{ 2 \over \gamma\sp {n-1}}\,. \label{vfp}
\end{equation}

\section{Soliton wave packets}
\setcounter{equation}{0}

The picture that emerges from our studies of one dimensional quantum
lattices with $f$ degrees of freedom and translational symmetry is as
follows.  For each value of the principle quantum number $n$ and wave
number $k$, there is a lowest energy eigenvalue.  These $f$ lowest
eigenvalues lie on a band (see Figure 1)
\begin{equation}
E = E_n(k)\,,\label{e1}
\end{equation}
where the wave number
\begin{equation}
k = {2\pi \nu \over f}\,,\label{k}
\end{equation}
and $\nu = 0, \pm1, \pm2, \ldots, \pm(f/2 -1), f/2$ for $f$ even and
$0, \pm1, \pm2, \ldots, \pm(f-1)/2$ for $f$ odd.  Each energy eigenvalue corresponds to a
pure eigenstate $|\psi_n(k)\rangle$, which is normalized as 
$\langle\psi_n(k)|\psi_n(k)\rangle = 1$.

For $\hat{H}$ being one of the Hamiltonian operators in Equations
(\ref{h1}), (\ref{h2}) or (\ref{h3}), solutions of the time dependent
Schr\"{o}dinger equation \begin{equation} {\rm i} {d{} \over
dt}|\Psi(t)\rangle = \hat{H}|\Psi(t)\rangle  \label{sch} \end{equation}
can be constructed as sums over the quantum number and the wave
number.  Thus for QDNLS and QAL
 \begin{equation}
|\Psi(t)\rangle = \sum_n a_n \sum_k G_n(k)|\psi_n(k)\rangle {\rm exp}\,
({\rm-i}E_n(k)t)\,, \label{Psi}
\end{equation}
where $n = 0,1,2,\ldots,\infty$, and $k$ takes the values between
$-\pi$ to $\pi$ that are indicated in Equation (\ref{k}).  Since
$\langle\Psi(t)|\Psi(t)\rangle = 1$, both the $a_n$ and the $G_n(k)$
are sets of complex numbers that satisfy the normalization conditions
\begin{equation}
\sum_n|a_n|\sp 2 = 1\,\;\;\;{\rm and}\;\;\;\sum_k|G_n(k)|\sp 2 = 1\,.  \label{nor3}
\end{equation}

It should be noted that Equation (\ref{Psi}) does not represent the
most general wave function that satisfies Equation (\ref{sch}) because
it is constructed only from eigenstates with eigenvalues on the soliton
bands.  It is seen from Figure 1 that the system has many other
eigenstates, and these are excluded from $|\Psi(t)\rangle$ as defined
in Equation (\ref{Psi}), which is characterized by two interdependent
properties:
\begin{itemize}

\item For given values of $n$, $k$, and the expansion coefficients, $|\Psi(t)\rangle$ has
the lowest energy, and 

\item Under the same conditions, $|\Psi(t)\rangle$ has the highest
probability of quanta being located near each other.

\end{itemize}
These properties are the basis for referring to $|\Psi(t)\rangle$ as a
``soliton wave packet''.

An effect of dispersion in the wave packet of Equation (\ref{Psi}) is
to introduce uncertainties in position $(j)$ and crystal momentum $(k)$
that satisfy the Heisenberg relation.  Such uncertainties appear in a
natural way because $G_n(k)$ is a discrete Fourier transform of the
soliton pulse shape so
\begin{equation}
\Delta k \times \Delta j \sim 1 \,. \label{ur}
\end{equation}
This point has been neglected at times in discussions of ``Davydov's
soliton'', which has been proposed as a polaronic means for transport
of energy or charge in protein \cite{sco92}.

In this section we present some properties of soliton wave packets.  In
addition to considering the differences between the QDNLS, QAL, and FP
models, we also distinguish between spectral problems, where $n$ has a
small, fixed value, and the correspondence limit for QDNLS and QAL,
where values of $n$ are large compared with unity and $|\Psi(t)\rangle$
is approximated by the solution of a classical nonlinear equation.

\subsection{Spectral problems }

The ground state ($n=0$) energy is zero for all three models, and the
corresponding eigenstate is $|\psi_0\rangle = [000\cdots0]$.  For $n=1$
the wave function of an exact eigenstate is given in Equation
(\ref{ps1}), again for all three models, and the energy eigenvalues are
given in Equation (\ref{b7}).  For $n=2$ exact expressions for
$|\psi_2(k)\rangle$ and $E_2(k)$ are presented in Section 3, and in
Section 4 we have derived approximate expressions for
$|\psi_n(k)\rangle$ and $E_n(k)$ that hold in the limits $\gamma \ll 1$
and $\gamma \gg 1$.

When the expression for $|\Psi(t)\rangle$ in Equation (\ref{Psi})
represents the wave function at a particular energy level (say
$n=n_0$), then
\begin{equation}
a_n = \delta_{n,n_0}\,.
\end{equation}
In measurements involving infra-red or Raman spectra on molecular
crystals, the wave length of the interacting radiation is much larger
than the dimensions of a unit cell so it is often reasonable to assume
$k \approx 0$, but one can arrange experiments for which this is not
the case.

The results of Section 3 permit exact calculations of features for the
three transitions:
\[
|\psi_0\rangle \rightarrow |\psi_1\rangle\,, 
\]
\[
|\psi_0\rangle \rightarrow |\psi_2\rangle\,,
\]
\[
|\psi_1\rangle \rightarrow |\psi_2\rangle\,.  
\]
{}From the results of Section 4, which
obtain in the limit $\gamma \gg 1$, approximate features can be
calculated for the transitions: 
\[
|\psi_0\rangle \rightarrow |\psi_n\rangle\,, 
\]
\[
|\psi_1\rangle \rightarrow |\psi_n\rangle\,,
\]
\[
 |\psi_2\rangle \rightarrow |\psi_n\rangle\,,
\]
\[
\cdots {\rm etc.}\cdots
\]
\[
|\psi_m\rangle \rightarrow |\psi_n\rangle\,.
\]

\subsection{ Large wave packets}

\vspace{.2in}
{\bf (i) QDNLS solitons in the Hartree approximation }
\vspace{.2in}

In the continuum limit this problem has been studied in detail by Lai
and Haus \cite{lh89} and by Wright \cite{wri91} as a model for the
propagation of solitons on an optical fiber.  The exact solution
exhibits two quantum effects: phase spreading, which is caused by
different values of $n$ in the wave packet, and dispersion, caused by
different values of $k$.  Lai and Haus have shown that the effects of
dispersion become negligible compared with phase spreading at large
values of the average number of quanta (bosons) in the soliton.

For the QDNLS equation with $n \gg 1$, the soliton wave function at a par\-ticu\-lar value of $n$ 
\begin{equation}
|\Psi_n(t)\rangle = \sum_k G_n(k)|\psi_n(k)\rangle{\rm exp}
[-{\rm i}E_n(k)t]
\end{equation}
is close to the Hartree approximation \cite{weh93}
\begin{equation}
|\psi_n(t)\rangle\sp {({\rm H})} = {1 \over
\sqrt{n!}}\left[\sum_{j=1}\sp f\Phi_{n,j}(t)b_j\sp {\dagger}\right]\sp n|0\rangle\;,\label{ha}
\end{equation}
where $\Phi_{n,j}(t)$ is a solution of
\begin{equation}
{\rm i}{d\Phi_{n,j}
\over dt} + (\Phi_{n,j+1} + \Phi_{n,j-1}) + \gamma
(n-1)|\Phi_{n,j}|\sp 2\Phi_{n,j}=0 \label{h}
\end{equation}
that satisfies the normalization condition
\begin{equation}
\sum_{j=1}\sp f|\Phi_{n,j}(t)|\sp 2 = 1\,. \label{ic}
\end{equation}

Since ${}\sp {(H)}\langle\psi_n(j,t)|b_j|\psi_n(j,t)\rangle\sp {(H)}=0
$, it is convenient to choose the $\{a_n\}$ in Equation (\ref{Psi}) so
that $|\Psi(t)\rangle$ is the coherent wave packet
\begin{equation}
|\Psi(t)\rangle = \sum_n {n_0\sp {n/2} \over \sqrt{n!}}{\rm exp}
\left(-{n_0 \over
2}\right)|\psi_n(j,t)\rangle\sp {(H)}\,. \label{Psi1}
\end{equation}

In the correspondence limit, $n_0 \gg 1$, 
\begin{equation}
\langle\Psi(t)|b_j|\Psi(t)\rangle \rightarrow A_j(t)\,,
\end{equation}
where $A_j(t)$ is a solution of the classical equation
\begin{equation}
{\rm i}{dA_j \over dt} + (A_{j+1} + A_{j-1}) + \gamma|A_j|\sp 2 A_j = 0 \label{a}
\end{equation}
with normalization
\begin{equation}
\sum_{j=1}\sp f |A_j|\sp 2 = n_0\,.
\end{equation}

\vspace{.1in}

For $\gamma > $ O(1) both classical solutions of Equation (\ref{a}) and
solutions of Equation (\ref{h}) do not propagate; they are pinned to
the lattice by the Peierls barrier \cite{fe91}.  The exact expression
for $|\Psi(t)\rangle$ is seen from Equation (\ref{v1}) to have a
maximum wave packet velocity
\begin{equation}
V_{\rm m} = {2n_0 \over (n_0 - 1)! \gamma\sp {n_0-1}}
\end{equation}
for $n_0 \gg 1$.  Thus the exact quantum mechanical solution also
becomes pinned in the correspondence limit in the sense that $V_{\rm m}
\rightarrow 0$ strongly as $n_0 \rightarrow \infty$.

\vspace{.2in}
{\bf (ii) QAL solitons }
\vspace{.2in}

For $\gamma \ll 1$ the QAL equation approaches the QDNLS equation and
the wave function of a QAL soliton is close to that given by Equations
(\ref{ha}) and (\ref{Psi1}).  For $\gamma > {\rm O}(1)$ there is---to
our knowledge---no Hartree approximation, and the two systems are quite
different.   From Equations (\ref{psie}) and (\ref{psio})
\begin{equation}
|\psi_n(k)\rangle  = { 1 \over \sqrt{f}}\sum_{j=1}\sp {f} ({\rm e}\sp
{{\rm i}k}\hat T)\sp {j-1}[n00\cdots0] + {\rm O}(\gamma\sp {-1}) 
\label{qal17}
\end{equation}
and
\begin{equation}
b_j\hat{T}\sp {j-1}[n00\cdots0] = \beta(n)\hat{T}\sp
{j-1}[(n-1)00\cdots0]\,,
\end{equation}
where $\beta(n)$ is defined in Equation (\ref{beta}).  Thus it is
possible to construct
\begin{equation}
|\Psi(t)\rangle = c\sum_{j=1}\sp f |\Psi_j(t)\rangle\,,
\end{equation}
where $|\Psi_j(t)\rangle$ is the coherent state
\begin{equation} 
|\Psi_j(t)\rangle \doteq \sum_{n=0}\sp {\infty}\left(
{2 \over \gamma}\right)\sp {n(n-1)/4}
\Phi_j\sp n(t)\hat{T}\sp {j-1}[n00\cdots0]\,,
\end{equation}
for which
\begin{equation}
b_j|\Psi_j(t)\rangle = \Phi_j(t)|\Psi_j(t)\rangle \label{bj}
\end{equation}
and 
\begin{equation}
c = \left[\sum_{n=0}\sp {\infty}\sum_{j=1}\sp f \left({|\Phi_j(0)|
\sp 2 \over
\beta(n)}\right)\sp n\right]\sp {-1/2}\,.
\end{equation}
Thus the Heisenberg operator equation
\begin{equation}
{\rm i}{db_j \over dt} + (b_{j+1} + b_{j-1})(1 + 
{\gamma \over 2}b_j\sp {\dagger}b_j) = 0\,,
\end{equation}
and Equation (\ref{bj}) imply that
\begin{equation}
{\rm i}{d\Phi_j \over dt} + (\Phi_{j+1} + \Phi_{j-1})(1 + {\gamma \over 2}|\Phi_j|\sp 2) = 0\,.
\label{al} 
\end{equation}

{}From Equation (\ref{ve}) we see that the maximum group velocity is
unbounded as $n \rightarrow \infty$.  This is consistent with the fact
that Equation (\ref{al}) supports moving solitons that do not become
pinned as $\gamma$ increases \cite{al76b}.

\vspace{.2in}
{\bf (iii) FP solitons}
\vspace{.2in}

Having arisen in the context of classical dynamics, the term
``soliton'' is essentially a classical concept, and a ``quantum
soliton'' is often considered to be an object that becomes a classical
soliton in the correspondence limit.  For the fermionic polaron model,
however, there is no correspondence limit because the number of
fermions is at most equal to the number of freedoms.  Nonetheless one
can construct a wave function of the form of Equation (\ref{Psi}),
where $n = 0,1,2,\ldots,f$.  Since this wave packet shares many
properties with Equation (\ref{Psi}), and the FP equation can be
analyzed using the quantum inverse scattering method \cite{pz86}, it is
appropriate to call $|\Psi(t)\rangle$ in Equation (\ref{Psi}) a quantum
soliton.

Since there is no correspondence limit for an FP soliton, it is not
inappropriate to choose a particular value for $n$ (say $n_0$) for the
sum in Equation (\ref{Psi}).  Thus $a_n = \delta_{n,n_0}$, and for
$\gamma \gg 1$ an approximate picture of our soliton (with velocity
$v$) becomes
\[
[0000 \cdots 0000\overbrace{\underbrace{11\cdots1}_{v\,\rightarrow}}\sp {n_0}0000 \cdots 0000]\,.
\]
However this diagram does not show uncertainties in position $(j)$ and
momentum $(k)$, which are required by the Heisenberg principle and
included in the structure of Equation (\ref{Psi}).  For $n=n_0$ this
equation takes the form
\begin{equation}
|\Psi_{n_0}(t)\rangle = \sum_k G_{n_0}(k)|\psi_{n_0}(k)\rangle 
{\rm exp}[-{\rm i}E_{n_0}(k)t]\,,
\label{Psi5} \end{equation}
where, from Equation (\ref{fpps}),
\begin{equation}
|\psi_{n_0}(k)\rangle \doteq { 1 \over \sqrt{f}} \sum_{j=1}\sp f 
({\rm e}\sp {{\rm i}k}
\hat{T})\sp {j-1}[11\cdots100\cdots0]\,, \end{equation}
and, from Equation (\ref{efp}),
\begin{equation}
E_{n_0}(k) \doteq -(n_0-1)\gamma - {2 \over \gamma\sp {n_0 -1}}
{\rm cos}k\,.
\end{equation}
Since $G_{n_0}(k)$ is the probability amplitude for the FP soliton to
have momentum $k$, Equation (\ref{Psi5}) can be written as
\begin{equation} |\Psi_{n_0}(t)\rangle \doteq { 1 \over
\sqrt{f}}\sum_{j=1}\sp f F_{n_0}(j,t)\hat{T}\sp {j-1}[11\cdots100
\cdots 0]\,, \end{equation}
where
\begin{equation}
F_{n_0}(j,t) \equiv \sum_k G_{n_0}(k) {\rm exp}[{\rm i}(k(j-1) - E_{n_0}(k)t)]\,. \label{f3}
\end{equation}
Thus uncertainties in position and momentum are related by Equation 
(\ref{ur}).  If $k$ lies
within the range
\begin{equation}
k = k_0 \pm {\Delta k \over 2}\,,
\end{equation}
then the soliton speed is in the range
\begin{equation}
v \pm {\Delta v \over 2} = {2 \over \gamma\sp {n_0 - 1}}
\left[{\rm sin}k\,{\rm cos}{\Delta k \over 2}
\pm {\rm cos}k\,{\rm sin}{\Delta k \over 2}\right]\,. 
\end{equation}

Since wave functions of an FP soliton are dominated by components of
the form
\[
[\cdots00\underbrace{11\cdots11}_{n\;{\rm times}}00\cdots]\,,
\]
it becomes more extended as the number of fermions increases.  For an
infinite number of freedoms on the lattice, one could allow the number
of fermions in Equation (\ref{Psi}) to grow without bound, but the size
of the soliton would also become infinite.
 
For $n=f$, $|\psi_f\rangle = [111\ldots1]$ with energy $E_f = -\gamma
f$, while for $f=0$, $|\psi_0\rangle = [000\ldots0]$ with energy
$E_0=0$.  In general the exact eigenfunctions for $n=f-m$ will have the
same structure as those for $n=m$, but with the 1's and 0's
interchanged.  Thus for $n = f-m$, where $m = 0,1,\ldots,f$, the
corresponding energy will be
\begin{equation}
E_{f-m}(k) = -\gamma(f-2m) + E_m(k)\,, \label{fp1}
\end{equation}
where $E_m(k)$ is the soliton energy band in the case $n=m$. 

\section{Conclusions}
\setcounter{equation}{0}

The primary aim of this work is to make clear the concept of a quantum
soliton by presenting several specific examples.  Lattice solitons are
of particular interest in this effort because it is possible---in
certain cases---to examine details of the corresponding wave functions
and to appreciate their complex character.  The properties of quantum
solitons are found to depend strongly upon  the level of anharmonicity
and the commutation relations that characterize the fundamental
quanta.

The number state method of analysis, which has been somewhat obscured
by the quantum inverse scattering method, is found to be a useful
computational tool for such problems as well as a helpful theoretical
perspective.

Whereas previous publications have concentrated on solving the quantum
problem with only a few quanta using the number state method, we have
shown that an approximate version of this method, valid in the limit of
large nonlinearity, is also useful for larger values of $n$. In
particular, we have found general expressions for the binding energy,
the effective mass and the maximum group velocity in this limit for
arbitrary $n>1$.

\vspace*{9pt}
\begin{tabular}{|l|c|c|c|} \hline
 model  &       $E_b$   &       $m\sp *$   &       $V_{\rm m}$ \\ 
\hline \hline
 QDNLS  &       $\frac{1}{2} n(n-1)\gamma$ & ${{(n-1)!}\over{2n}}
\gamma\sp {n-1}$
        &       $ {2n}\over{(n-1)!\gamma\sp {n-1}}$ \\ \hline
 QAL    &       $2 {\left( {\gamma}\over 2 \right)}\sp {(n-1)/2}$
        &       $ {{n\sp 2}\over 2}\left({2 \over {\gamma}}\right)
\sp {(n-1)/2}$
        &       $ {2 \over n} \sin({\pi\over n})
                \left( {\gamma \over 2}\right)\sp {(n-1)/2}$\\ \hline
 FP     &       $ (n-1)\gamma$          &    ${\gamma\sp {n-1}}\over 2$
        &       $ {2 \over {\gamma\sp {n-1}}}$\\ \hline     
\end{tabular}

\vspace*{9pt}

We note that the results for the group velocities confirm that the
QDNLS and the FP solitons get pinned for large $\gamma$ whereas this is
not the case for QAL solitons, in agreement with the findings of
classical soliton theory.

This method can also be applied to Hubbard models \cite{mrr90}, which
are of interest as theories of superconductivity and ferromagnetism.
This application will be discussed elsewhere.

Finally we note that the band structures shown in Figure 1 may be
somewhat misleading because they are for two fundamental quanta
$(n=2)$.  With $n\geq3$, exploratory numerical and theoretical studies
show that additional bands reside in the gap between the lowest
(soliton) band and the principal continuum band.  The significance of
these bands will be the subject of future research.

\section*{Acknowledgements}

We acknowledge support from the British Council, the SERC Nonlinear
Systems Initiative, the EC under SCI-0229-C89-100079/JU1 and from the
NSF under Grant No. DMS-9114503.  One of us (JCE) is grateful to Jack
Carr for useful conversations.

\end{document}